\documentclass[12pt]{iopart}
\usepackage{geometry}
\usepackage{units}
\usepackage{graphicx}

\begin{document}

\title{Tip-surface interactions in dynamic atomic force microscopy}
\author{Daniel Platz$^1$, Daniel Forchheimer$^1$, Erik A. Thol\'{e}n$^2$ and David B. Haviland$^1$}
\address{$^1$Royal Institute of Technology (KTH), Section for Nanostructure Physics,
Albanova University Center, SE-106 91 Stockholm, Sweden}
\address{$^2$Intermodulation Products AB, Vasavägen 29, SE-169 58 Solna, Sweden}
\ead{platz@kth.se}

\begin{abstract}
In atomic force microscopy (AFM) tip-surface interactions are usually
considered as functions of the tip position only, so-called force
curves. However, tip-surface interactions often depend on the tip
velocity and the past tip trajectory. Here, we introduce a compact
and general description of these interactions appropriate to dynamic
AFM where the measurement of force is restricted to a narrow frequency
band. We represent the tip-surface interaction in terms of a force
disk in the phase space of position and velocity. Determination of
the amplitude dependence of tip-surface forces at a fixed static probe
height allows for a comprehensive treatment of conservative and dissipative
interactions. We illuminate the fundamental limitations of force reconstruction
with narrow band dynamic AFM and we show how the amplitude dependence
of the Fourier component of the force at the tip oscillation frequency,
gives qualitative insight into the detailed nature of the tip-surface
interaction. With minimal assumptions this amplitude dependence force
spectroscopy allows for a quantitative reconstruction of the effective
conservative tip-surface force as well as a position-dependent damping
factor. We demonstrate this reconstruction on simulated intermodulation
AFM data.
\end{abstract}

\noindent{\it Keywords\/}: atomic force microscopy, measurement of force, mechanical 
resonators, MEMS/NEMS, dissipation, intermodulation

\pacs{03.65.Ge, 07.79.Lh, 07.10.Pz, 62.20.-F, 85.85.+j}

\maketitle

\section{Introduction}

For the understanding of surface reactions and the characterization
of materials it is desirable to measure local forces close to a sample
surface. The most common method to measure these surface forces is
atomic force microscopy (AFM)\cite{Binnig1986}. Historically, the
first force measurements were static measurements for which the force
is presented as a scalar function of the static tip-sample separation,
the so-called force curve\cite{Burnham1993,Butt2005b}. This representation
is sufficient for conservative forces but the total tip-surface force
may also contain contributions from dissipative forces. Since dissipative
forces depend on probe velocity and past trajectory, dynamic force
spectroscopy methods are required for their measurement. Moreover,
the visualization of dissipative forces as a function of position
is valid only for a specific probe trajectory and simple force curves
cannot capture the full character of the interaction. Despite the
development of several dynamic methods\cite{Durig2000,Durig2000a,Stark2002,Sader2005,Lee2006,Holscher2006,Hu2008,Katan2009}
surface forces are still usually treated as functions of the probe
position only and represented by simple force curves.

Here, we present a comprehensive framework for the representation
and analysis of complex surface forces as they are measured by dynamic
AFM. We concentrate on the most common modes of dynamic AFM: amplitude-modulated
AFM (AM-AFM) and frequency-modulated AFM (FM-AFM), which can be considered
as narrow frequency band methods\cite{Platz2012b}. We explore the
fundamental limit of force reconstruction with narrow band dynamic
AFM at fixed probe height and show how minimal assumptions allow for
a quantitative reconstruction of the tip-surface interaction.

\subsection{Cantilever-based dynamic force measurements}

At the heart of the AFM apparatus is a micro-cantilever with a sharp
tip. The cantilever is firmly clamped at one end and the tip is located
at the other end which can move freely. It is assumed that surface
forces only act on the tip whereas the rest of the cantilever does
not experience significant surface forces. In dynamic AFM an additional
external drive force is applied to maintain an oscillatory motion.
Thus, the dynamics are governed by the force between tip and surface,
the external drive force and the properties of the cantilever beam.

Since the cantilever is a three dimensional continuum object its motion
is usually described by the amplitudes of different oscillation eigenmodes.
In general, these modes can cause the cantilever to bend in all directions
in space. However, the cantilever is positioned such that the softest
flexural modes bend the beam in a plane orthogonal to the surface
plane. We restrict ourselves to the case where only these flexural
modes are excited by the drive force. Due to this experimental configuration
the cantilever is much more susceptible to the component of the tip-surface
force which is orthogonal to the surface plane. This component of
the force is typically the most dominant component and the influence
of lateral force components is considered negligible. In this case
the cantilever acts as a mechanical projector which reacts only to
one component of a three dimensional force vector field.

The deflection $w$ of a cantilever of length $L$ orthogonal to surface
is described by a one dimensional Euler-Bernoulli equation\cite{Landau1968}
\begin{equation}
EI\frac{\partial^{4}w(x,t)}{\partial x^{4}}+\mu\frac{\partial^{2}w(x,t)}{\partial t^{2}}=F\left(x,w(x),t\right)\label{eq:euler-bernoulli}
\end{equation}
where $E$ is the Young's modulus, $I$ is the second moment of area,
$\mu$ is the mass per unit length of the cantilever, $x$ is the
position coordinate along the cantilever beam and $t$ is the time
variable. The force term $F$ includes the surface forces acting as
a point-like load at position $x=L$, the external drive force and
the hydrodynamic damping due to the surrounding medium\cite{Sader1998}.
To express the solution to equation (\ref{eq:euler-bernoulli}) in
terms of the oscillation eigenmodes, one performs a separation of
variables and obtains a solution for $w(x,t)$ that is a linear combination
of different mode shapes $\Phi^{(n)}(x)$ with time-dependent amplitudes
$q^{(n)}(t)$,
\begin{equation}
w(x,t)=\sum_{n=1}^{\infty}q^{(n)}(t)\Phi^{(n)}(x)
\end{equation}
where the mathematically orthogonal modes $\Phi^{(n)}$ have different
resonance frequencies $\omega_{0}^{(n)}$. 

In most dynamic AFM modes only the first eigenmode is externally excited.
The spectrum of the resulting tip motion is then confined in a narrow
frequency band around the first flexural resonance frequency $\omega_{0}^{(1)}$\cite{Platz2012b}.
Due to the fact that the resonance frequencies of higher eigenmodes
are not integer multiples of the first resonance frequency $\omega_{0}^{(1)}$
and that the eigenmodes have high quality factors, only the first
mode contributes to the cantilever motion and the tip motion $z(t)=w(x=L,t)+h$
can be approximated as
\begin{equation}
z(t)\approx q^{(1)}(t)\Phi^{(1)}(L)+h
\end{equation}
where $h$ is the static probe height above the surface. The time-dependence
is given by an effective harmonic oscillator equation\cite{Stark2000a,Melcher2007}
\begin{equation}
\ddot{z}+\frac{\omega_{0}^{(1)}}{Q^{(1)}}\dot{z}+k_{\mathrm{c}}^{(1)}(z-h)=F_{\mathrm{drive}}(t)+F_{\mathrm{ts}}\left(z,\dot{z},\{z(t)\}\right)\label{eq:ho-em}
\end{equation}
where the dot denotes differentiation with respect to the time $t$,
$Q^{(1)}$ is the quality factor of the first flexural resonance,
$k_{\mathrm{c}}^{(1)}$ the effective mode stiffness, $F_{\mathrm{drive}}$
the time-dependent external drive force and $F_{\mathrm{ts}}$ the
force between tip and surface. Thus, equation (\ref{eq:ho-em}) reduces
the basic physics of dynamic AFM to that of a damped harmonic oscillator
which is moving in an nonlinear force field $F_{\mathrm{ts}}$ and
is subject to a time-dependent drive force $F_{\mathrm{drive}}$. 

Simulations starting from a given tip-surface force reveal multiple
oscillation states\cite{Garcia1999,Garcia2000} and period multiplications\cite{Hashemi2008},
qualitative features of the dynamics which have also been observed
in experiments\cite{SanPaulo2000,Hu2006,Jamitzky2006}. However, the
fundamental challenge in AFM is actually the inverse problem: Given
an accurate and well-calibrated measurement of the dynamics $z(t)$,
how can we quantitatively determine the tip-surface force. Different
sophisticated methods have been developed to solve this inverse problem
in the framework of the single harmonic oscillator model\cite{Durig2000,Durig2000a,Stark2002,Sader2005,Lee2006,Holscher2006,Hu2008,Katan2009}
but they usually assume simple forms of the tip-surface force $F_{\mathrm{ts}}$
in which the interaction depends on the instantaneous tip position
only.

\section{The force disk}

\subsection{General properties of surface forces}

Surface forces depending on the instantaneous tip position and velocity
can be represented as two dimensional function in the $z$-$\dot{z}$
phase plane. The representation of a force depending on the history
of the tip motion $\left\{ z(t)\right\} $, for example hysteretic
forces due to capillary formation or chemical bonds, is more difficult.
However, for high quality factor oscillators with large stored energy,
the motion is restricted to a narrow band near resonance and the steady
state motion is well approximated by a sinusoidal trajectory. For
a sinusoidal drive signal at the first resonance frequency the motion
is then given by\cite{Paulo2002} 
\begin{equation}
z(t)=A\cos(\omega_{0}^{(1)}t+\phi)+h\label{eq:tip-motion}
\end{equation}
where $A$ is the oscillation amplitude, $\phi$ is the phase lag
with respect to the drive. The motion orbits defined by equation (\ref{eq:tip-motion})
do not intersect each other in the $z$-$\dot{z}$ phase plane and
thus every point in the phase plane can be mapped to a unique tip
trajectory. We assume that the force along each trajectory does not
depend on previous oscillation cycles. In this manner we can incorporate
the dependence of the force on the past tip trajectory into the dependence
on the instantaneous tip position and velocity.

\begin{figure}[th]
\begin{center}
\includegraphics{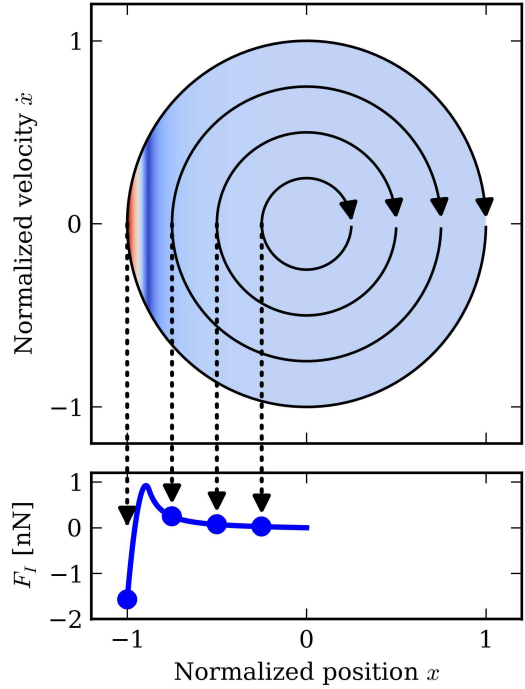}
\end{center}
\caption{Force disk in the $x$-$\dot{x}$ plane of a purely conservative force.
For typical values of the static probe height and the oscillation
amplitude the interaction is very localized at the surface. The depicted
orbits correspond to purely sinusoidal tip motion. The orbits do not
intersect each other so every point in the plane can be mapped to
a unique oscillation amplitude. There is one value of $F_{I}$ and
$F_{Q}$ for each orbit which we assign to the lower turning point.\label{fig:force-disk}}
\end{figure}

In every experiment there exists a maximum oscillation amplitude $A_{\mathrm{max}}$
and velocity $v_{\mathrm{max}}$. We normalize the tip position and
velocity to the maximum amplitude and velocity and subtract the static
probe height,
\begin{eqnarray}
x & \equiv & \frac{z-h}{A_{\mathrm{max}}}\\
\dot{x} & \equiv & \frac{\dot{z}}{v_{\mathrm{max}}}
\end{eqnarray}
to obtain a new force function on the closed unit disk in the $x$-$\dot{x}$
plane,
\begin{equation}
F_{\mathrm{ts}}^{(x)}(x,\dot{x})=F_{\mathrm{ts}}(A_{\mathrm{max}}x,v_{\mathrm{max}}\dot{x}).
\end{equation}
The motion defined by equation (\ref{eq:tip-motion}) corresponds
to circular orbits in the $x$-$\dot{x}$ plane with a maximum radius
of 1 as shown in fig. \ref{fig:force-disk}. Thus, it is sufficient
to define the force function $F_{ts}^{(x)}$ as a function on the
closed unit disk in the $x$-$\dot{x}$ plane. This force disk is
a compact and general description of the interaction between the tip
and the surface in narrow band dynamic AFM.

One regularly distinguishes between conservative and non-conservative
or dissipative surface forces. The force during a complete tip oscillation
cycle can readily be decomposed into an effective conservative and
an effective non-conservative force by finding the symmetric and anti-symmetric
part of the measured force around the lower turning point of the tip
motion\cite{Durig2000a,Sader2005}. This decomposition ensures that
the energy dissipation integral 
\begin{equation}
E_{\mathrm{dis}}=\oint_{C[z(t)]}F_{\mathrm{ts}}\ dz=\int_{0}^{T}F_{\mathrm{ts}}\left(z(t),\dot{z}(t),\left\{ z(t)\right\} \right)\dot{z}(t)\ dt\label{eq:E-dis}
\end{equation}
equals zero for the effective conservative force. In contrast, dissipative
forces are responsible for the energy dissipation. One should note
that forces of different physical origin can contribute to the total
surface force. Each of the effective forces may therefore contain
contributions from forces which alone would be of purely conservative,
purely dissipative or of mixed character.

In the same manner we are able to decompose the force disk into an
effective conservative disk $F_{\mathrm{c}}^{(x)}$ and into an effective
dissipative disk $F_{\mathrm{nc}}^{(x)}$,
\begin{equation}
F_{\mathrm{ts}}^{(x)}(x,\dot{x})=F_{\mathrm{c}}^{(x)}(x,\dot{x})+F_{\mathrm{nc}}^{(x)}(x,\dot{x}),
\end{equation}
by requiring that for the conservative force disk, no energy is dissipated
during one complete oscillation cycle of the sinusoidal tip motion.
This implies that the conservative force disk is symmetric with respect
to the x-axis whereas the non-conservative force disk is anti-symmetric\cite{Hu2008}
\begin{eqnarray}
F_{\mathrm{c}}(x,-\dot{x}) & = & F_{\mathrm{c}}(x,\dot{x})\label{eq:fc-sym}\\
F_{\mathrm{nc}}(x,-\dot{x}) & = & -F_{\mathrm{nc}}(x,\dot{x})\label{eq:fnc-anti-sym}
\end{eqnarray}

Functions on the unit disk are naturally expressed in polar coordinates
so we perform a change of variables to obtain a force function $F_{\mathrm{ts}}^{(r)}$
that depends on the normalized amplitude $r$ and the instantaneous
phase of the oscillation $\theta$,
\begin{equation}
F_{\mathrm{ts}}^{(x)}(x,\dot{x})\longrightarrow F_{\mathrm{ts}}^{(r)}(r,\theta).
\end{equation}
Similar to functions on the unit sphere $F_{\mathrm{ts}}^{(r)}$ can
be expanded into a set of orthogonal functions such that 
\begin{equation}
F_{\mathrm{ts}}^{(r)}(r,\varphi)=\sum_{n=0}^{\infty}\sum_{m=-n}^{n}a_{n}^{(m)}Z_{n}^{(m)}(r,\theta).\label{eq:zernike-expansion}
\end{equation}
A natural choice of the basis functions $Z_{n}^{(m)}$ are the Zernike
polynomials\cite{Zernike1934} which can be defined as
\begin{equation}
\label{eq:zernike-definition}
Z_n^{(m)}=\cases{
R_{n}^{(|m|)}(r)\cos(m\theta)  &for $m\geq0$ \\
R_{n}^{(|m|)}(r)\sin(m\theta)  &for $m<0$\\
}
\end{equation}
where the polynomials $R_{n}^{(|m|)}$ are given by\cite{PrataJr.1989}
\begin{equation}
R_{n}^{(|m|)}(r) =
\cases{
\sum_{k=0}^{(n-|m|)/2}\frac{(-1)^{k}(n-k)!}{k!((n+m)/2-k)!((n-m)/2-k)!}r^{n-2k} &for $m-n\ \mathrm{even}$\\
0 &for $m-n\ \mathrm{odd}$\\
}
\end{equation}
 and fulfill the orthogonality relation
\begin{equation}
\int_{0}^{1}R_{n}^{(|m|)}(r)R_{n'}^{(|m|)}(r)r\ dr=\frac{1}{2(n+1)}\delta_{n,n'}.\label{eq:r-orthogonality}
\end{equation}
We will use this Zernike expansion of the force disk in the following
section to investigate which parts of the force disk are measurable
with narrow band dynamic AFM.

\subsection{Probing the force disk with dynamic narrow frequency band AFM}

In dynamic AFM the tip-surface force can also be considered as a time-dependent
force acting on the oscillating tip. In AM-AFM and FM-AFM only the
Fourier component of the time-dependent tip-surface force at the tip
oscillation frequency is measurable above the noise floor. Higher
frequency components of the force are filtered out by the high-quality
factor resonance of the cantilever. The force Fourier component at
the oscillation frequency can be expressed as a real-valued component
$F_{I}$ that is in-phase with a sinusoidal tip motion and a real-valued
component $F_{Q}$ that is quadrature to the tip motion\cite{Platz2012b}.
With equation (\ref{eq:tip-motion}) we can write $F_{I}$ and $F_{Q}$
as two integral equations 
\begin{eqnarray}
F_{I} & = & \frac{1}{T}\int_{0}^{T}F_{\mathrm{ts}}\left(A\cos\left(\omega_{0}^{(1)}t\right)+h,-\omega_{0}^{(1)}A\sin\left(\omega_{0}^{(1)}t\right)\right)\cos\left(\omega_{0}^{(1)}t\right)dt\label{eq:fi-int}\\
F_{Q} & = & \frac{1}{T}\int_{0}^{T}F_{\mathrm{ts}}\left(A\cos\left(\omega_{0}^{(1)}t\right)+h,-\omega_{0}^{(1)}A\sin\left(\omega_{0}^{(1)}t\right)\right)\sin\left(\omega_{0}^{(1)}t\right)dt\label{eq:fq-int}
\end{eqnarray}
The component $F_{I}$ is the so-called virial of the tip motion which
is only affected by the effective conservative force\cite{Paulo2002}.
In contrast, $F_{Q}$ is connected to the dissipative interaction
and comparison with the energy dissipation integral in equation (\ref{eq:E-dis})
yields
\begin{equation}
E_{\mathrm{dis}}=-2\pi AF_{Q}.\label{eq:fq-edis}
\end{equation}
Through their dependence on the tip motion $z(t)$ and the tip velocity
$\dot{z}(t)$ the force components $F_{I}$ and $F_{Q}$ are functions
of the oscillation amplitude $A$ and the static probe height $h$.
Alternatively, $F_{I}$ and $F_{Q}$ can be considered as functions
of $h$ and the lower turning point $z_{\mathrm{min}}$ such that
at fixed static probe height, the force disk and the $F_{I}(A)$ and
$F_{Q}(A)$ curves share the same position axis as shown in figure \ref{fig:force-disk}. 

While imaging with conventional dynamic AFM the feedback is working
to keep the the oscillation amplitude constant. Thus, only one value
for $F_{I}$ and $F_{Q}$ is measured. Therefore, the combination
of imaging and force measurement is not possible in conventional dynamic
AFM and most force spectroscopy techniques rely on a measurement of
the $h$ dependence of $F_{I}$ and $F_{Q}$. As an alternative approach,
we recently introduced the rapid measurement of the oscillation amplitude
dependence of $F_{I}$ and $F_{Q}$ at fixed static probe height with
Intermodulation AFM\cite{Platz2012b}. To understand what information
about the force disk can be extracted from a measurement of $F_{I}(A)$
and $F_{Q}(A)$ we insert the Zernike expansion of the tip-surface
force disk, equation (\ref{eq:zernike-expansion}), into the integral
equations (\ref{eq:fi-int}) and (\ref{eq:fq-int})

\begin{eqnarray}
F_{I}(A) & = & \int_{0}^{2\pi}\sum_{n=0}^{\infty}\sum_{m=0}^{n}a_{n}^{(m)}R_{n}^{(m)}(\nicefrac{A}{A_{\mathrm{max}}})\cos(m\theta)\cos(\theta)d\theta\nonumber \\
 & = & \pi\sum_{n=0}^{\infty}a_{n}^{(1)}R_{n}^{(1)}(\nicefrac{A}{A_{\mathrm{max}}}),\label{eq:fi-zernike}\\
F_{Q}(A) & = & \int_{0}^{2\pi}\sum_{n=0}^{\infty}\sum_{m=1}^{n}a_{n}^{(m)}R_{n}^{(m)}(\nicefrac{A}{A_{\mathrm{max}}})\sin(m\theta)\sin(\theta)d\theta\nonumber \\
 & = & \pi\sum_{n=0}^{\infty}a_{n}^{(-1)}R_{n}^{(-1)}(\nicefrac{A}{A_{\mathrm{max}}}).\label{eq:fq-zernike}
\end{eqnarray}
We note that $F_{I}(A)$ and $F_{Q}(A)$ only depend on the coefficients
$a_{n}^{(\pm1)}$ with $m=\pm1$ where the individual coefficients
can be recovered by using the orthogonality relation (\ref{eq:r-orthogonality}).
This implies that the measurable information in narrow band dynamic
AFM is fundamentally limited. It is not possible to reconstruct an
arbitrary force disk from the experimentally available quantities
$F_{I}(A)$ and $F_{Q}(A)$. Nevertheless, the $F_{I}(A)$ and $F_{Q}(A)$
curves provide qualitative insight into the interaction between tip
and surface.

The fundamental limitation at fixed static probe height is due to
the limited frequency band. To increase the measurable information
additional frequency bands have to be considered. One way to achieve
this is to externally excite multiple cantilever eigenmodes\cite{Rodrguez2004,Kawai2009,Solares2010}.
In liquid environments higher eigenmodes exhibit measurable response
even in the absence of an external drive, due to the mode's low quality
factor\cite{Melcher2008a,Melcher2009}. However, multiple eigenmodes
require a higher dimensional phase space description of the dynamics
and the representation of force must be appropriately adapted. Another
way to introduce additional frequency components is the measurement
of higher harmonics of the tip motion, although higher harmonics are
only measurable under special conditions\cite{Stark2002,Legleiter2006}
or with specialized cantilevers\cite{Sahin2007}.

\subsection{Examples of force disks\label{sub:force-disk-examples}}

\begin{figure}[th]
\begin{center}
\includegraphics{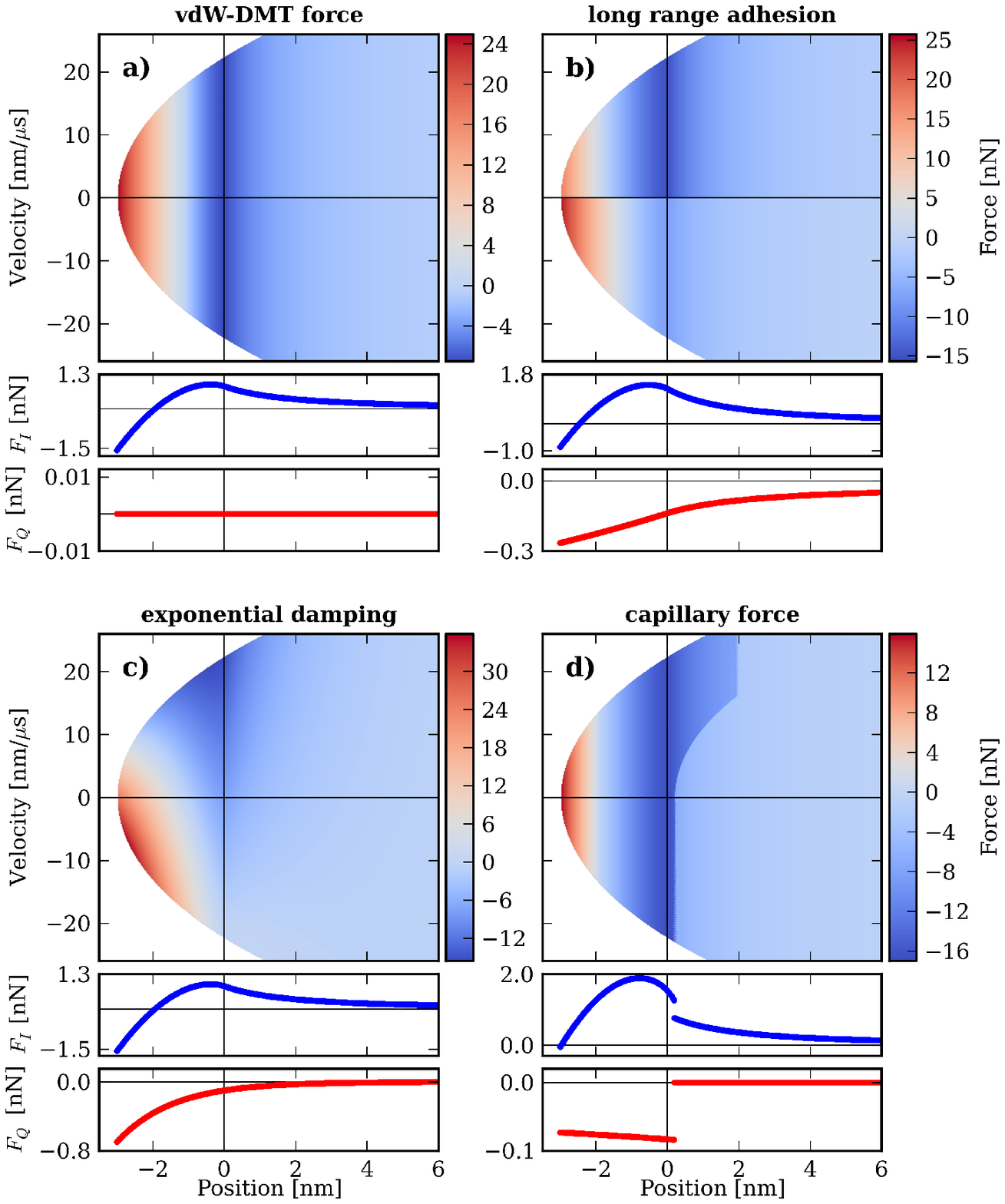}
\end{center}
\caption{Model force disks in the $z$-$\dot{z}$ plane at a static probe height
of $h=22\ \mathrm{nm}$ and a maximum oscillation amplitude $A_{\mathrm{max}}=25\ \mathrm{nm}$.
The models all build on a purely conservative vdW-DMT force (a) and
add long range adhesion(b), position-dependent viscous damping(c)
and capillary interactions(d) respectively. We concentrate on the
region close to the surface and display the disk sections together
with the corresponding $F_{I}$ and $F_{Q}$ curves as functions of
the lower motion turning point.
\label{fig:zooms-disk}}
\end{figure}

In figure \ref{fig:zooms-disk} four examples of the force disk for
physically relevant model forces are shown in the $z$-$\dot{z}$
plane. Since the interaction is very localized we focus on to the
region close to the surface. 
In figure \ref{fig:zooms-disk}a a conservative van der Waals-Derjagin-Muller-Toropov
(vdW-DMT) force is shown which is given by equation (\ref{eq:f-vdw-dmt})\cite{DERJAGUIN1975}
. The symmetry of the force disk with respect to the velocity and
the fact that $F_{Q}$ equals zero for all motion turning points,
is due to the vdW-DMT force being purely conservative. From the $F_{I}(A)$
curve we can conclude that the conservative force can be decomposed
in two regions of different character. Further away from the surface
$F_{I}$ is positive which corresponds to an average attractive force
acting on the tip. When the tip makes contact with the surface $F_{I}$
quickly becomes negative due to the rapid turn-on of the strongly
repulsive force.

Figure \ref{fig:zooms-disk}b shows a force disk that models a long-range
dissipative force\cite{Garcia2006a}. The model builds on a conservative
vdw-DMT force with a Hamaker constant whose value is different for
the tip approach and retract. This discontinuous behaviour is defined
by equation (\ref{eq:f-long}) and appears in the force disk as a
discontinuous asymmetry with respect to the velocity. The force shown
in figure \ref{fig:zooms-disk}b illustrates how the addition of a so-called
dissipative interaction can modify the effective conservative tip-surface
force. Despite the fact that the vdw-DMT parameters are the same as
for figure \ref{fig:zooms-disk}a, the $F_{I}$ curves differ for the
two models. The added ``dissipative'' interaction actually contains
an effective conservative component, which makes the total effective
conservative force more attractive and less repulsive. The nonzero
values of $F_{Q}$ indicate the presence of a dissipative force which
is already present relatively far away from the surface. Moreover,
from the shape of the $F_{Q}$ curve we can draw conclusions about
the nature of the dissipative interaction: Further away from the surface
$F_{Q}$ is inversely proportional to the squared oscillation amplitude.
Below the contact point at $z=0$, the derivative of $F_{Q}$ with
respect to the oscillation amplitude reveals that $F_{Q}$ is inversely
proportional to the oscillation amplitude in this region. This behaviour
indicates that the energy dissipated during each oscillation cycle
grows linearly with oscillation amplitude from the contact point on.

In contrast to figure \ref{fig:zooms-disk}b the force displayed in
figure \ref{fig:zooms-disk}c is an example of an additional dissipative
interaction that does not modify the effective conservative force.
The force is defined in equation (\ref{eq:f-exp}) and combines a
vdw-DMT force with a position-dependent viscous damping\cite{Gotsmann1999}.
Due to the linear dependence of the force on the tip velocity the
$F_{I}$ curve in fig \ref{fig:zooms-disk}c does not differ from
the $F_{I}$ curve for the purely conservative vdW-DMT force in figure
\ref{fig:zooms-disk}a . The damping coefficient for the force in
figure \ref{fig:zooms-disk}c depends exponentially on position, which
results in a dissipated energy that is proportional to the product
of the oscillation amplitude and modified Bessel function of first
kind in the oscillation amplitude\cite{Gotsmann1999}. Hence, the
$F_{Q}$ curve also follows a modified Bessel function of first kind. 

Physically different interactions are encountered for capillary surface
forces which can be due to adsorbed water layers on the surface. The
common model is based on a vdW-DMT force with an additional hysteretic
adhesion force that turns on and off instantaneously when the tip
passes certain threshold positions as defined in equation (\ref{eq:f-zitzler})\cite{Zitzler2002}.
Up to a threshold position $z_{\mathrm{on}}$ the force disk is symmetric
with respect to velocity. At the threshold position the tip makes
contact with the water layer on the surface which results in an additional
attractive force. Due to this adhesion a capillary neck builds up
between the tip and the surface when the tip retracts again. This
necks breaks at a position $z_{\mathrm{off}}$ that is different from
$z_{\mathrm{on}}$. The water neck effectively extends the region
of strong attraction during the tip retract. The abrupt force jumps
are also visible in the $F_{I}(A)$ and $F_{Q}(A)$ curves. Since
the dissipated energy is independent of the oscillation amplitude,
the $F_{Q}(A)$ curve is inversely proportional to the oscillation
amplitude which becomes more pronounced in the derivative of $F_{Q}$
with respect to the oscillation amplitude.

For all example surface forces considered here the $F_{I}(A)$ and
$F_{Q}(A)$ curves give insight into the character of effective conservative
and dissipative interactions between the tip and the surface. With
ImAFM\cite{Platz2008,Platz2010} these curves are measured at each
image point while scanning at imaging speeds\cite{Platz2012b}. This
rapid and information-rich data acquisition technique allows for an
enhanced interpretation of imaging contrast, a polynomial reconstruction
of the tip-surface force\cite{Hutter2010a,Platz2012a} and the extraction
of material properties\cite{Forchheimer2012}.

\section{Amplitude-dependence force spectroscopy (ADFS)}

The ultimate goal of dynamic AFM is the combination of high-speed
and high-resolution imaging with high accuracy force measurements.
However, imaging is performed at static probe height $h$ above the
surface and we have seen that in the case of fixed $h$ it is not
possible to reconstruct a complete force disk from the available $F_{I}(A)$
and $F_{Q}(A)$ curves (see equations (\ref{eq:fi-zernike})
and (\ref{eq:fq-zernike})). Often we can make physically well-motivated
assumptions about the tip-surface interaction which effectively correlate
the Zernike expansion coefficients of the force disk. Under these
assumptions a measurement of the amplitude dependence of $F_{I}$
and $F_{Q}$ is sufficient for a quantitative reconstruction
the tip-surface interaction. We call this method amplitude dependence
force spectroscopy (ADFS) and in this section we show the reconstruction
of the effective conservative tip-surface interaction as well as the
reconstruction of a position-dependent viscous damping.

\subsection{Reconstruction of conservative tip-surface interactions}

In most cases it can be assumed that the effective conservative tip-surface
force does not depend on tip velocity and tip motion history but is
rather a function of the tip position only,
\begin{eqnarray}
F_{\mathrm{c}}(z,\dot{z}) & = & F_{\mathrm{c}}(z).\label{eq:ft-c-z}\\
\Rightarrow F_{\mathrm{c}}^{(x)}(x,\dot{x}) & = & F_{\mathrm{c}}^{(x)}(x)\label{eq:f-c-x}
\end{eqnarray}
We recently demonstrated how this assumption allows for the reconstruction
of the effective conservative surface force $F_{c}$ with ADFS\cite{Platz2013}.
Using the equations (\ref{eq:f-c-x}) and (\ref{eq:fc-sym}) we rewrite
the integral equation (\ref{eq:fi-int}) for $F_{I}$ 

\begin{eqnarray}
F_{I}(A) & = & \frac{1}{2\pi}\int_{0}^{2\pi}F_{\mathrm{c}}\left(A\cos\theta+h\right)\cos(\theta)d\theta\\
 & \stackrel{z\equiv A\cos\theta}{=} & \frac{1}{\pi}\int_{-A}^{A}F_{\mathrm{c}}(z+h)\frac{z/A}{\sqrt{A^{2}-z^{2}}}dz.
\end{eqnarray}
Usually, the interaction length of the force is small (several nm)
compared to the oscillation range (tens of nm). So we can neglect
the upper half of the integration interval,
\begin{eqnarray}
F_{I}(A) & \approx & \int_{-A}^{0}F_{\mathrm{c}}(z+h)\frac{z/A}{\sqrt{A^{2}-z^{2}}}dz\\
 & \stackrel{u\equiv z^{2}}{=} & \frac{1}{2\pi A}\int_{A^{2}}^{0}\frac{F_{\mathrm{c}}\left(-\sqrt{u}+h\right)}{\sqrt{A^{2}-u}}du\label{eq:fi-int-u}
\end{eqnarray}
With the definitions
\begin{eqnarray}
\tilde{A} & \equiv & A^{2}\label{eq:A2}\\
\tilde{F}_{\mathrm{c}}(u) & \equiv & F_{\mathrm{c}}(-\sqrt{u}+h)\\
\tilde{F}_{I}(\tilde{A}) & \equiv & -2\pi\sqrt{\tilde{A}}F_{I}(\sqrt{\tilde{A}})
\end{eqnarray}
equation (\ref{eq:fi-int-u}) becomes
\begin{equation}
\tilde{F}_{I}(\tilde{A})=\int_{0}^{\tilde{A}}\frac{\tilde{F}_{\mathrm{c}}(u)}{\sqrt{\tilde{A}-u}}du.\label{eq:f_i_abel_int}
\end{equation}
We note that $\tilde{F}_{I}(\tilde{A})$ in equation (\ref{eq:f_i_abel_int})
is the Abel transform of the force $\tilde{F}_{\mathrm{c}}(u)$. The
Abel transform has a unique inverse\cite{Arfken2005,Groetsch2007}
with which we solve equation (\ref{eq:f_i_abel_int}) for the force
$F_{\mathrm{c}}(-z+h)$
\begin{eqnarray}
\tilde{F}_{\mathrm{c}}(u) & = & \frac{1}{\pi}\frac{d}{du}\int_{0}^{u}\frac{\tilde{F}_{I}(\tilde{A})}{\sqrt{u-\tilde{A}}}d\tilde{A}\\
\Rightarrow F_{\mathrm{c}}(-z+h) & = & -\frac{1}{z}\frac{d}{dz}\int_{0}^{z^{2}}\frac{\sqrt{\tilde{A}}F_{I}(\sqrt{\tilde{A}})}{\sqrt{z^{2}-\tilde{A}}}d\tilde{A}.\label{eq:force_abel}
\end{eqnarray}
In practice, the integral in equation (\ref{eq:force_abel}) has to
be evaluated numerically which is made difficult by the square root
singularity at the upper integration limit. It is therefore advantageous
to perform the substitution $y^{2}=z^{2}-\tilde{A}$ in the integral
to remove the singularity and to improve the numerical stability,
\begin{eqnarray}
F_{\mathrm{c}}(-z) & \stackrel{y^{2}\equiv z^{2}-\tilde{A}}{=} & -\frac{2}{z}\frac{d}{dz}\int_{0}^{z}\sqrt{z^{2}-y^{2}}F_{I}\left(\sqrt{z^{2}-y^{2}}\right)dy.
\end{eqnarray}
The resulting integral is then suitable for standard numerical integration
methods like simple trapezoidal integration.

\subsection{Reconstruction of position-dependent damping}

Often dissipation is modeled as a position-dependent damping or friction
coefficient $\lambda(z)$ such that the dissipative force is given
by
\begin{equation}
F_{\mathrm{nc}}(z,\dot{z})=\lambda(z)\dot{z}.\label{eq:f_pos_damping}
\end{equation}
In this case we can reconstruct the effective damping function $\lambda(z)$
from the amplitude-dependence of $F_{Q}(A)$. We start with rewriting
the integral equation (\ref{eq:fq-int}) by using the equations (\ref{eq:fnc-anti-sym})
and (\ref{eq:f_pos_damping}).
\begin{eqnarray}
F_{Q}(A) & = & -\frac{\omega A}{2\pi}\int_{0}^{2\pi}\lambda(A\cos\theta+h)\sin^{2}\theta\ d\theta\\
 & \stackrel{z\equiv A\cos\theta}{=} & -\frac{\omega}{\pi A}\int_{-A}^{A}\lambda(z+h)\sqrt{A^{2}-z^{2}}\ dz
\end{eqnarray}
As for the conservative force we assume that the oscillation range
is bigger than the interaction range of the dissipative force such
that $\lambda(z)=0$ for $z\geq0$,
\begin{eqnarray}
F_{Q}(A) & \approx & -\frac{\omega}{\pi A}\int_{-A}^{0}\lambda(z+h)\sqrt{A^{2}-z^{2}}\ dz\\
 & \stackrel{u=z^{2}}{=} & -\frac{\omega}{2\pi A}\int_{0}^{A^{2}}\frac{\lambda\left(-\sqrt{u}+h\right)}{\sqrt{u}}\sqrt{A^{2}-u}\ du
\end{eqnarray}
We use definition (\ref{eq:A2}) and define additionally 
\begin{eqnarray}
\tilde{F}_{Q}(\tilde{A}) & \equiv & -\frac{2\pi\sqrt{\tilde{A}}}{\omega}F_{Q}\left(\sqrt{\tilde{A}}\right)\\
\tilde{\lambda}(u) & \equiv & \frac{\lambda\left(-\sqrt{u}+h\right)}{\sqrt{u}}
\end{eqnarray}
to arrive at
\begin{equation}
\tilde{F}_{Q}(\tilde{A})=\int_{0}^{\tilde{A}}\tilde{\lambda}(u)\sqrt{\tilde{A}-u}\ du.\label{eq:fq-conv}
\end{equation}
The integral (\ref{eq:fq-conv}) represents a convolution of $\tilde{\lambda}(u)$
with $\sqrt{u}$ which can readily be solved in Laplace space (see
\ref{sec:Inversion-damping-int}). For tip motion as defined by equation
(\ref{eq:tip-motion}) the position-dependent damping coefficient
is then given by 
\begin{eqnarray}
\tilde{\lambda}(\tilde{A}) & = & \frac{2}{\pi}\frac{d^{2}}{d\tilde{A}^{2}}\int_{0}^{\tilde{A}}\frac{\tilde{F}_{Q}(u)}{\sqrt{\tilde{A}-u}}\ du\\
\Rightarrow\lambda\left(-\sqrt{\tilde{A}}+h\right) & = & -\frac{4\sqrt{\tilde{A}}}{\omega_{0,1}}\frac{d^{2}}{d\tilde{A}^{2}}\int_{0}^{\tilde{A}}\frac{\sqrt{u}F_{Q}(\sqrt{u})}{\sqrt{\tilde{A}-u}}\label{eq:gamma-abel}
\end{eqnarray}

\subsection{Numerical results}

\begin{figure}[th]
\begin{centering}
\includegraphics{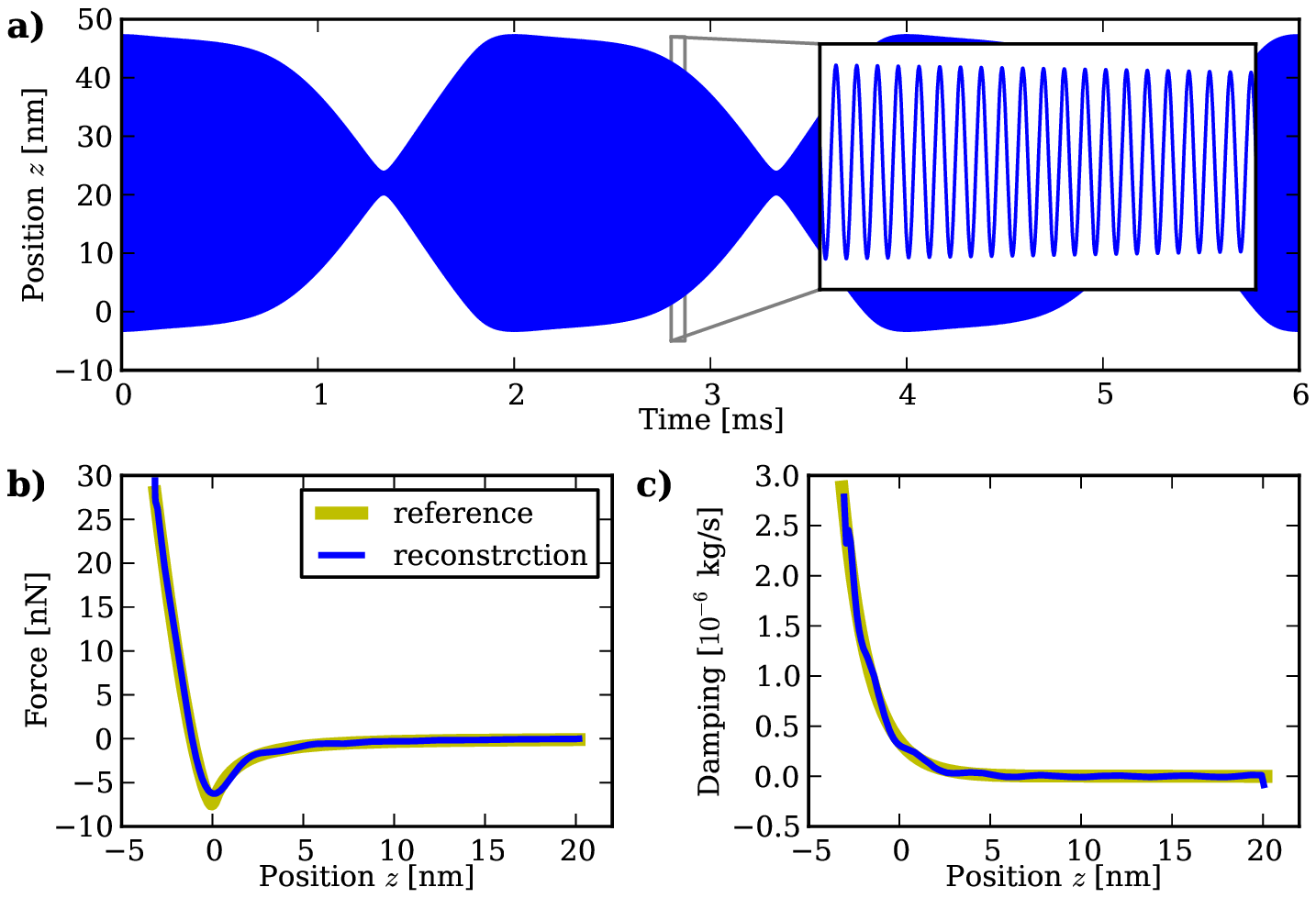}
\end{centering}
\caption{Simulated tip motion in the time domain(a). On a fast time scale the
tip motion is nearly purely sinusoidal with an amplitude that is modulated
on a much slower time scale. The ADFS reconstructions of the effective
conservative force curve(a) and the position-dependent damping curve(b)
are in excellent agreement with the reference curves used in the simulation.
\label{fig:adfs-results}}
\end{figure}

To validate the reconstruction of the effective conservative tip-surface
force and the position dependent-damping we simulate the tip motion
in a scalar force field with a vdW-DMT component and exponential damping
as defined by equation (\ref{eq:f-exp}). To integrate the equation
of motion (\ref{eq:ho-em}) we use the CVODE solver with adaptive
step-size and discrete event detection\cite{Hindmarsh2005} to account
for the piecewise definition of the tip-surface force in equation
(\ref{eq:f-exp}). We assume a cantilever typically used for experiments
under ambient conditions with a quality factor of $Q^{(1)}=400.0$,
a stiffness of $k_{c}^{(1)}=40.0\ \mathrm{N/m}$ and a resonance frequency
of $f_{0}^{(1)}=300.0\ \mathrm{kHz}$. To probe the tip-surface interaction
we use an ImAFM drive scheme such that the free oscillation amplitude
is modulated between 0 and 35 nm in a time window of $T=2\ \mathrm{ms}$
at a fixed static probe height of $h=22.0\ \mathrm{nm}$ above the
sample surface as shown in figure \ref{fig:adfs-results}a.

From the tip motion close to the surface the $F_{I}(A)$ and $F_{Q}(A)$
curves are obtained\cite{Platz2012b} which are then used for the
numerical evaluation of the equations (\ref{eq:force_abel}) and (\ref{eq:gamma-abel}).
The results are shown in figure \ref{fig:adfs-results}.
Both the reconstructed effective conservative force and the position-dependent
damping are in excellent agreement with the force and damping actually
used in the simulation. For the conservative force the actual and
the reconstructed force deviate by at most 4.5 \% of the maximum force.
This deviation is the result of the rapid change in the force around
the contact point, which generate Fourier components with amplitudes
typically below the noise level in AFM systems and these components
have been neglected in the reconstruction.

We note that the interaction is localized in the region between -3.2
nm and 7.5 nm. A more optimized drive scheme would therefore modulate
the oscillation amplitude only in this interaction region, thereby
sampling the interaction more smoothly since more turning points lie
in the interaction region during the same measurement time\cite{Arlett2011,
Boisen2011}.

\section{Conclusions}

We have introduced a comprehensive framework for the representation
of tip-surface interaction at fixed static probe height in terms of
a ``force disk''. This framework allows for a complete description
of forces depending on the instantaneous tip position and velocity,
appropriate to narrow band dynamic AFM. We show that it not possible
to fully reconstruct arbitrary tip-surface interactions from narrow
band dynamic AFM measurements. However, by considering different tip
oscillation amplitudes, the experimental curves $F_{I}(A)$ and $F_{Q}(A)$
give insight in the character of the effective conservative and the
effective dissipative interaction. With minimal additional assumptions
these curves can be used for quantitative reconstructions of the effective
conservative tip-surface force and the effective position-dependent
damping.

The framework introduced here gives a more solid theoretical foundation
for the visualization and analysis of tip-surface forces than traditional
force curves. Furthermore, we expect that the notion of a force disk
will inspire new multifrequency force probing schemes expanding the
capabilities of conventional narrow band dynamic AFM. On the experimental
side amplitude-dependence force spectroscopy enables rapid force measurements
while scanning sample surfaces at normal imaging speeds. From the
obtained force volume data sets different mechanical properties of
the sample surface can be derived without the constraints and limitations
of over-simplified interaction models. 

Since the underlying physics of the theory presented here is the classical
harmonic oscillator moving in an external force field. We see wide-spread
applications of this method of analysis in other fields using resonant
detection techniques based on different kinds of high-quality factor
resonators\cite{Arlett2011, Boisen2011}.

\ack{}{}

The authors acknowledge financial support from the Knut and Alice
Wallenberg Foundation, Swedish Research Council (VR) and the Swedish
Government Agency for Innovation Systems (VINNOVA).

\appendix

\section{Definition of model surface forces}

In section \ref{sub:force-disk-examples} we introduced various force
models of which the mathematical definitions are given here. Also
given are the values of the parameters used when plotting or simulating
with these models.

The basis of all models forms a vdW-DMT force which is defined as
\begin{equation}
F_{\mathrm{DMT}}(z) = 
\cases{
-\frac{HR}{6(a_{0}-z)^{2}} &for $z\geq0$\\
-\frac{HR}{6a_{0}^{2}}+\frac{4}{3}E^{*}\sqrt{R}(-z)^{3/2} &for $z<0$\\ 
}
\label{eq:f-vdw-dmt}
\end{equation}
where $H=3.28\cdot10^{-17}\ \mathrm{J}$ is the Hamker constant, $R=10.0\ \mathrm{nm}$
is the tip radius, $a_{0}=2.7\ \mathrm{nm}$ is the intermolecular
distance and $E^{*}=1.5\ \mathrm{GPa}$ is the effective stiffness
of the tip-surface system.

The long-range dissipative force is defined as 
\begin{equation}
F_{\mathrm{long}}(z,\dot{z})  =  
\cases{
-\frac{H_{\mathrm{app}}R}{6(a_{0}-z)^{2}}                                   &for $z\geq0\ \mathrm{and}\ \dot{z}\leq0$\\
-\frac{H_{\mathrm{app}}R}{6a_{0}^{2}}+\frac{4}{3}E^{*}\sqrt{R}(-z)^{3/2}    &for $z<0\ \mathrm{and}\ \dot{z}\leq0$ \\
-\frac{H_{\mathrm{ret}}R}{6(a_{0}-z)^{2}}                                   &for $z\geq0\ \mathrm{and}\ \dot{z}>0$\\
-\frac{H_{\mathrm{ret}}R}{6a_{0}^{2}}+\frac{4}{3}E^{*}\sqrt{R}(-z)^{3/2}    &for $z<0\ \mathrm{and}\ \dot{z}>0$\\
}
\label{eq:f-long}
\end{equation}
which extends the vdW-DMT model by a Hamaker constant that depends
on if the tip is approaching the surface ($H_{\mathrm{app}}=H$) or
is retracting from it ($H_{\mathrm{ret}}=2H$)

The vdW-DMT with an additional viscous damping term depending exponentially
on position is given by
\begin{equation}
F_{\mathrm{exp}}(z,\dot{z})  =  
\cases{
-\frac{HR}{6(a_{0}-z)^{2}}-\gamma_{0}\exp\left(-\nicefrac{z}{z_{\gamma}}\right)\dot{z}                                 &for $z\geq0$\\
-\frac{HR}{6a_{0}^{2}}+\frac{4}{3}E^{*}\sqrt{R}(-z)^{3/2}-\gamma_{0}\exp\left(-\nicefrac{z}{z_{\gamma}}\right)\dot{z}  &for $z<0$\\
}
\label{eq:f-exp}
\end{equation}
where $\gamma_{0}=3.5\ \mathrm{Js}$ is the damping factor and $z_{\gamma}=1.5\ \mathrm{nm}$
is the damping decay length.

Capillary interactions are modeled as
\begin{equation}
F_{\mathrm{cap}}(z,\left\{ z(t)\right\} )  = 
\cases{
-\frac{HR}{6(a_{0}-z)^{2}}                                                                                &for $z\geq z_{\mathrm{off}}$\\
-\frac{HR}{6(a_{0}-z)^{2}}                                                                                &for $z\geq z_{\mathrm{on}}\ \mathrm{and}\ z<z_{\mathrm{off}}\ \mathrm{and}\ m=0$ \\
-\frac{HR}{6(a_{0}-z)^{2}}-\frac{4\pi\gamma_{\mathrm{H_{2}O}}R}{1+\frac{z}{h_{\mathrm{H_{2}O}}}}          &for $z\geq z_{\mathrm{on}}\ \mathrm{and}\ z<z_{\mathrm{off}}\ \mathrm{and}\ m=1$ \\
-\frac{HR}{6(a_{0}-z)^{2}}-\frac{4\pi\gamma_{\mathrm{H_{2}O}}R}{1+\frac{z}{h_{\mathrm{H_{2}O}}}}          &for $z<z_{\mathrm{on}}\ \mathrm{and}\ z\geq0$ \\
-\frac{HR}{6a_{0}^{2}}-\frac{4\pi\gamma_{\mathrm{H_{2}O}}R}{1+\frac{a_{0}}{h_{\mathrm{H_{2}O}}}}+\frac{4}{3}E^{*}\sqrt{R}(-z)^{3/2}   &for $z\leq0$\\
}
\label{eq:f-zitzler}
\end{equation}
where $\gamma_{\mathrm{H_{2}O}}=72\cdot10^{-3}\ \mathrm{J/m^{2}}$
is the surface energy of water and $h_{\mathrm{H_{2}O}}=0.1\ \mathrm{nm}$
is the effective thickness of the adsorbed water layer. The state
variable $m$ is set to 1 when the tip makes contact with the water
layer at $z_{\mathrm{on}}=0.2\ \mathrm{nm}$ and is set to 0 when
the water neck breaks at $z_{\mathrm{off}}=1.94\ \mathrm{nm}$. In
contrast to the model introduced in\cite{Zitzler2002} we assume that
the interface between the surface and the water layer is located at
$z=0\ \mathrm{nm}$.

\section{Inversion of the damping integral\label{sec:Inversion-damping-int}}

The reconstruction of the position-dependent damping $\lambda(z)$
from $\hat{F}_{Q}(A)$ requires in the inversion of the integral equation
\begin{equation}
\tilde{F}_{Q}(\tilde{A})=\int_{0}^{\tilde{A}}\tilde{\lambda}(u)\sqrt{\tilde{A}-u}\ du
\end{equation}
We study this convolution in Laplace space where it becomes
\begin{equation}
\mathcal{L}\left\{ \tilde{F}_{Q}(\tilde{A})\right\} =\mathcal{L}\left\{ \int_{0}^{\tilde{A}}\tilde{\lambda}(u)\sqrt{\tilde{A}-u}\ du\right\} =\mathcal{L}\left\{ \tilde{\lambda}(\tilde{A})\right\} \mathcal{L}\left\{ \sqrt{\tilde{A}}\right\} .
\end{equation}
The Laplace transform of the square root is $\mathcal{L}\left\{ \sqrt{\tilde{A}}\right\} =\frac{\sqrt{\pi}}{2}s^{-3/2}$
which yields
\begin{eqnarray}
\frac{1}{s^{2}}\mathcal{L}\left\{ \tilde{\lambda}(\tilde{A})\right\}  & = & \frac{2}{\pi}s^{-1/2}\mathcal{L}\left\{ \tilde{F}_{Q}(\tilde{A})\right\} \\
 & = & \frac{2}{\pi}\mathcal{L}\left\{ \frac{1}{\sqrt{\tilde{A}}}\right\} \mathcal{L}\left\{ \tilde{F}_{Q}(\tilde{A})\right\} \\
 & = & \frac{2}{\pi}\mathcal{L}\left\{ \int_{0}^{\tilde{A}}\frac{\tilde{F}_{Q}(u)}{\sqrt{\tilde{A}-u}}\ du\right\} 
\end{eqnarray}
Now, we use the relation between integration in the time domain and
its Laplace transform and obtain
\begin{equation}
\int_{0}^{\tilde{A}}d\tilde{A}_{1}\int_{0}^{\tilde{A}_{1}}d\tilde{A}_{2}\tilde{\lambda}(\tilde{A}_{2})=\frac{2}{\pi}\int_{0}^{\tilde{A}}\frac{\tilde{F}_{Q}(u)}{\sqrt{\tilde{A}-u}}\ du
\end{equation}
which we can solve for the damping function $\tilde{\lambda}(\tilde{A})$,
\begin{equation}
\tilde{\lambda}(\tilde{A})=\frac{2}{\pi}\frac{d^{2}}{d\tilde{A}^{2}}\int_{0}^{\tilde{A}}\frac{\tilde{F}_{Q}(u)}{\sqrt{\tilde{A}-u}}\ du
\end{equation}

\providecommand{\newblock}{}

\end{document}